\documentclass{article}
\usepackage{graphicx}
\usepackage[english]{babel}
\newcommand{\refeq}[1]{Eq.~(\ref{#1})}

\title{
\includegraphics[width=0.35\textwidth, bb=0 0 945 955]{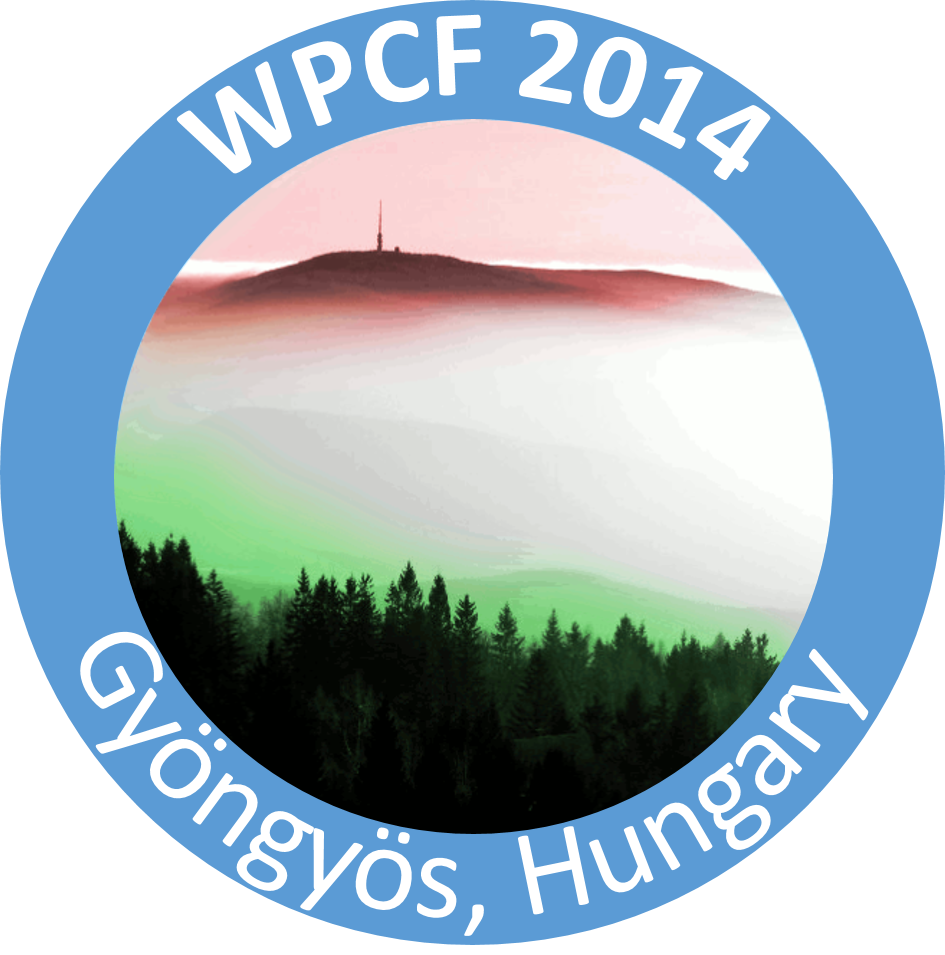}\\[1cm]
Event geometrical anisotropy and fluctuation viewed by HBT interferometry}
\author{{Takafumi Niida}\\[1ex]
University of Tsukuba\\ 
1-1-1 Tennoudai, Tsukuba, Ibaraki 305-8571, Japan\\
}

\begin{document}

\fontfamily{lmss}\selectfont
\maketitle

\begin{abstract}
Azimuthal angle dependence of the pion source radii was measured applying the event shape selection at the PHENIX experiment.
The measured final source eccentricity is found to be enhanced when selecting events with higher magnitude of the second-order flow vector, as well as the elliptic flow coefficient $v_{2}$.
The spatial twist of the particle-emitting source was also explored using a transport model. Results indicate a possible twisted source in the final state due to the initial longitudinal fluctuations.
\end{abstract}

\section{Introduction}
Higher-order flow coefficients $v_n$ are useful observables to constrain the properties of the quark-gluon plasma, such as a shear viscosity over entropy density ratio, in heavy ion collisions~\cite{chvn_phnx,CGCvsVn}. 
The $v_n$ are found to largely fluctuate in each event even if the collision centrality is fixed~\cite{ebeVn_atlas}, 
which is due to the initial spatial fluctuation of participating nucleons.
To control such event-by-event initial fluctuations, the event shape engineering was suggested~\cite{ESE}.
The event shape engineering can be performed by selecting the magnitude of the event flow vector $Q_{2}$:
\begin{eqnarray}
Q_2 &=& \sqrt{q_{2,x}^{2}+q_{2,y}^{2}}/\sqrt{M}, \\ \label{eq_qv}
q_{2,x} &=& \sum_i w_{i} \cos(2\phi_i), \,\,\,\,\, q_{2,y} = \sum_i w_{i} \sin(2\phi_i), \nonumber
\end{eqnarray}
where $M$ is the particle multiplicity, $\phi_i$ is the azimuthal angle of particles, and $w_i$ is a weight for particle $i$. 
The magnitude of $Q_2$ is proportional to the strength of event-by-event $v_2$ and also reflects the resolution of the second order event plane.
This technique could allow us to test the effect of the initial geometry on the final state distribution of emitted particles, 
which help to understand the medium response through the system evolution.

The event shape engineering focuses on the fluctuations in the transverse plane. 
It is naturally considered that there would be the fluctuations not only in the transverse plane but also in the longitudinal direction.
The presence of such fluctuations could cause a twisted source along the longitudinal direction~\cite{Bozek}.
The number of participants going to the forward and backward directions would be different, 
and therefore participant planes might also be different. 
As a result, the event plane angle in the final state could be different between the forward and backward angles.
Thus the initial spatial twist may survive in the final state as well as a twisted flow~\cite{ETW1,ETW2}. 

In this proceedings, we present results on HBT measurements using charged pions and applying the event shape engineering technique for Au+Au collisions at $\sqrt{s_{_{NN}}}$=200 GeV recorded with the PHENIX experiment.
Also, we examine the possibility of a spatially twisted source in the final state using HBT interferometry in AMPT model.

\section{The event shape engineering at the PHENIX}
The flow vector was determined by the Reaction Plane Detector (RXN, $1<|\eta|<2.8$).
In our analysis, the $w_i$ in~\refeq{eq_qv} reflects the multiplicity or energy measured in a segment $i$ of the RXN and M is $\sum w_i$.
The tracking for charged particles was performed by the Drift Chamber and the Pad Chamber, and the particle identification was done 
by the electromagnetic calorimeter, where they have the acceptance of $|\eta|<0.35$ and $|\phi|<\pi/2$ in the west and east central arms.

\begin{figure}[htb]
\begin{center}
\includegraphics[width=0.90\textwidth, clip, trim=0 0 0 0]{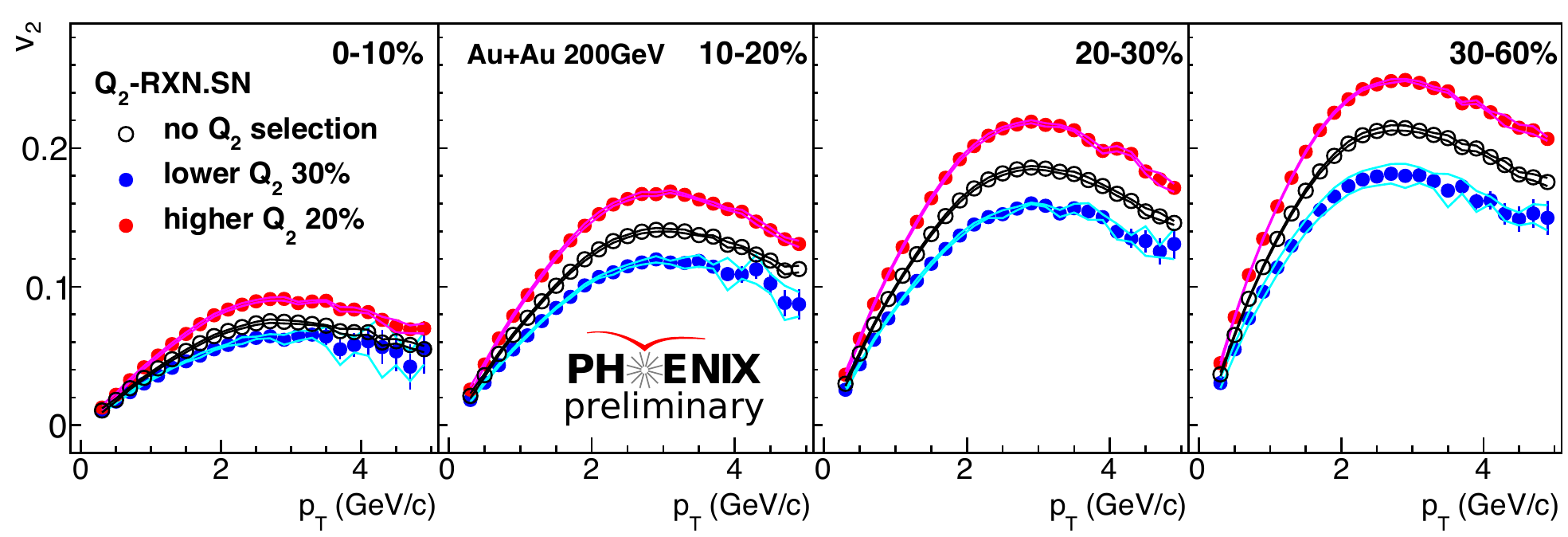}
\caption{\label{fig1}Charged hadron $v_2$ as a function of $p_{\rm T}$ for four centrally bins with and without the $Q_2$ selection in Au+Au 200 GeV collisions. The lower and higher $Q_{2}$ events were selected.}
\end{center}
\end{figure}

The measured $Q_2$ distributions were fitted with the Bessel-Gaussian function~\cite{ESE}, and then the events were classified with the magnitude of $Q_2$. Figure~\ref{fig1} shows charged hadron $v_2$ for the higher 20\% and the lower 30\% $Q_2$ events. Results without $Q_2$ selection are also plotted for the comparison.
%Fig1------------------------------------------------------------------------------
The effect of $Q_2$ selection on the $v_2$ is clearly observed, that is, the higher (lower) $Q_2$ enhances (decreases) the strength of $v_2$ compared to the $v_2$ without the $Q_2$ selection.

\begin{figure}[bht]
\begin{minipage}{0.6\textwidth}
\includegraphics[width=\textwidth,trim=0 10 0 0]{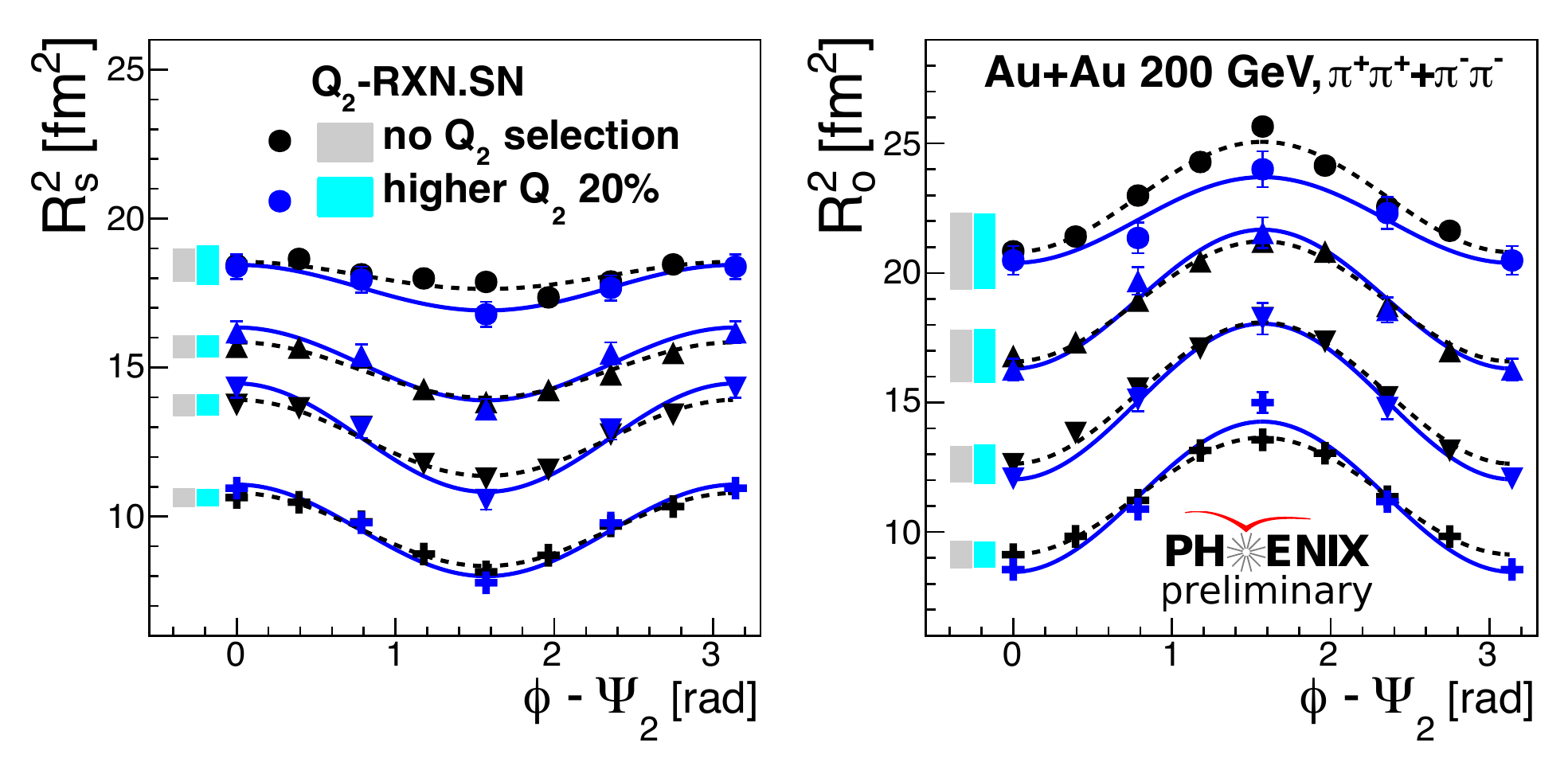}
\end{minipage}\hspace{0.5pc}%
\begin{minipage}{0.4\textwidth}
\includegraphics[width=0.85\textwidth,clip,trim=0 10 5 0]{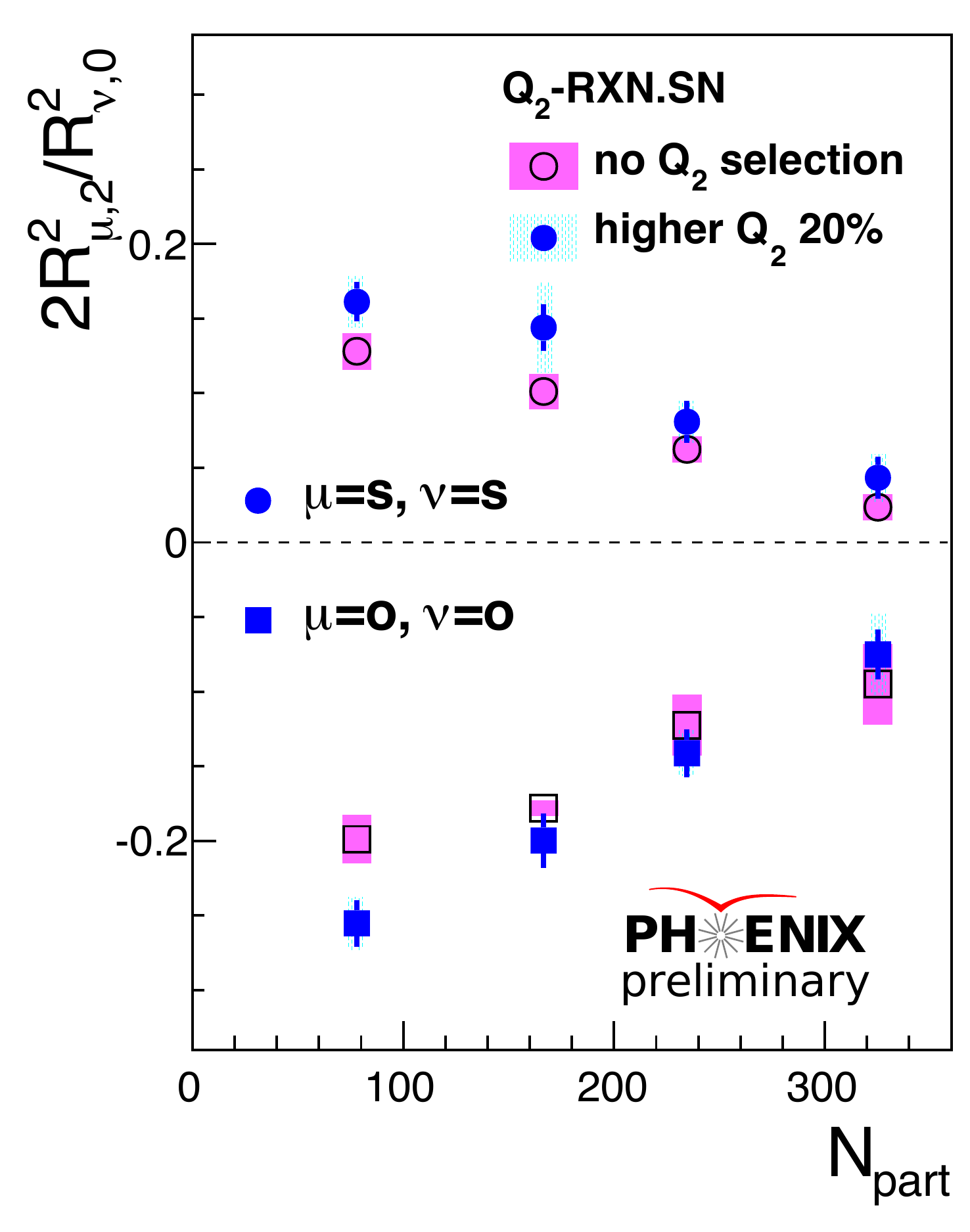}
\end{minipage} 
\caption{\label{fig2} (Left and center) Azimuthal angle dependence of $R_{\rm s}^{2}$ and $R_{\rm o}^{2}$ relative to $\Psi_2$. (Right) 2$^{\rm nd}$-order azimuthal oscillation on $R_{s}^{2}$ and $R_{o}^{2}$. In both panels, results with and without higher $Q_2$ selection are shown.}
\end{figure}

Then the $Q_2$ selection was applied to the HBT measurements using charged pion pairs. 
In the HBT analysis, pion pairs were analyzed with the out-side-long parameterization~\cite{OSL1,OSL2} in the longitudinally co-moving system. The effect of the event plane resolution was also corrected for both cases with and without $Q_2$ selection.
Figure~\ref{fig2}(left) shows the extracted pion HBT radii, $R_{\rm s}^2$ and $R_{\rm o}^2$, as a function of azimuthal pair angle $\phi$ relative to the second-order event plane $\Psi_2$. Results show that the higher $Q_2$ selection increases the oscillation strength compared to the case without $Q_2$ selection. 

These oscillations of HBT radii are supposed to be sensitive to the final source eccentricity at freeze-out, $\varepsilon_{\rm final}$. Blast-wave studies suggest that the quantity of $2R_{\rm s,2}^{2}/R_{\rm s,0}$ would be a good probe of $\varepsilon_{\rm final}$ in the limit of $k_{\rm T}=0$, where $k_{\rm T}$ denotes a mean transverse momentum of pairs. The oscillation amplitudes of $R_{\rm s}^2$ and $R_{\rm o}^2$ in a form of the final eccentricity are plotted as a function of the number of participants calculated by Glauber model in Fig.~\ref{fig2}(right). The higher $Q_2$ selection enhances the measured $\varepsilon_{\rm final}$ as well as $v_2$. 
It could be originating from a larger initial eccentricity, although there should be a bias from selecting the larger $v_2$ events  
because the radii modulations also depend on the anisotropy in the momentum space~\cite{bw,psi3}.
%Fig2------------------------------------------------------------------------------

\section{Study on the event twist effect}
The data of Pb+Pb 2.76 TeV collisions simulated using a Multi Phase Transport Model (AMPT, v2.25 with string melting)
were used for the study on the twist effect of the particle-emitting source in the final state. The impact parameter was fixed to $8$ fm.
For the HBT study, the interference effect between two identical particles, 
$1+\cos(\Delta {\bf r} \cdot \Delta {\bf p})$, was calculated and weighted to the relative pair momentum distributions. Then the correlation functions were reconstructed by taking a ratio of the distributions with and without the weight. Also, all charged pions were allowed to make a pair with each other including $\pi^+\pi^-$ to increase the statistics (the consistency between results for positive and negative pairs was checked).
The event plane was determined using particles in $4<|\eta|<6$, where particles were divided into two sub-groups.
A set of forward and backward event planes ($\Psi_2^{\rm F}, \Psi_2^{\rm B}$) were used for the event cut which requires finite difference between $\Psi_2^{\rm F}$ and $\Psi_2^{\rm B}$ as a event twist selection, whereas the other set of event planes was used for a reference angle of azimuthal HBT measurements. The effect of the event plane resolution was not taken into account in this study, assuming the resolution hardly affects the phase of the radii oscillations but the magnitude of the oscillations.

\begin{figure}[b]
\begin{center}
\includegraphics[width=0.92\textwidth,trim=0 20 0 20]{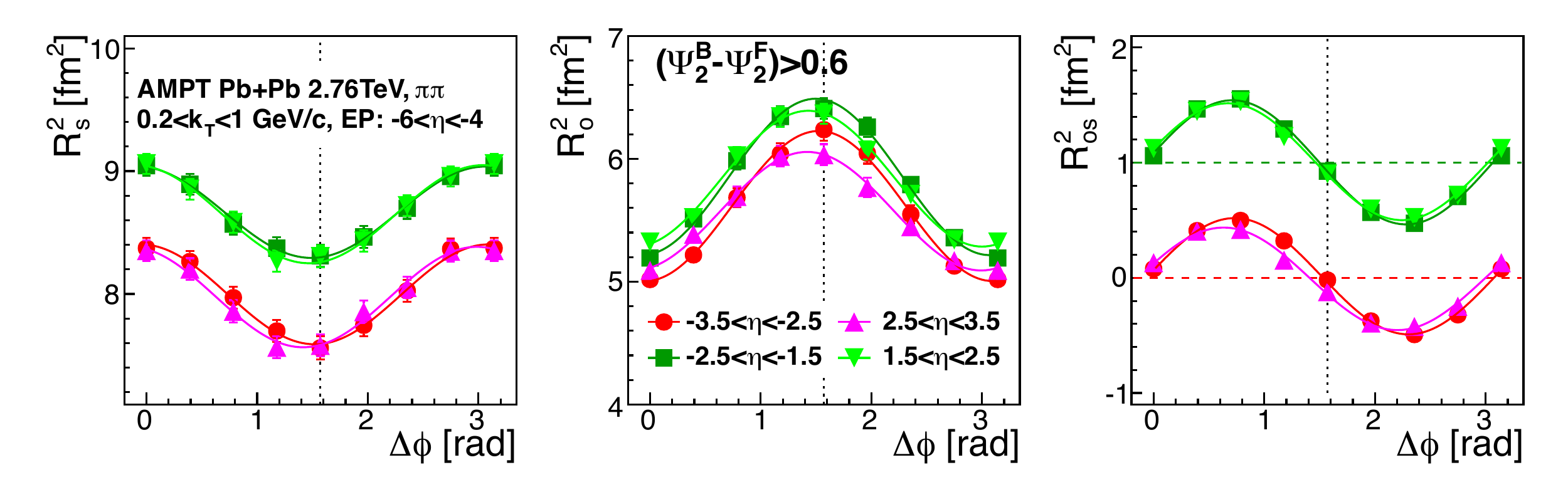}
\caption{\label{fig3} Azimuthal angle dependence of $R_{\rm s}^{2}$, $R_{\rm o}^{2}$, and $R_{os}^2$ relative to the backward $\Psi_2$ with the event cut of $(\Psi_2^{\rm B}-\Psi_2^{\rm F})>0.6$, where the dashed lines show $R_{\rm os}^2=0$ and the dotted lines show $\Delta\phi=\pi/2$. }
\end{center}
\end{figure}

\begin{figure}[tb]
\begin{center}
\includegraphics[width=0.9\textwidth,trim=0 0 0 10]{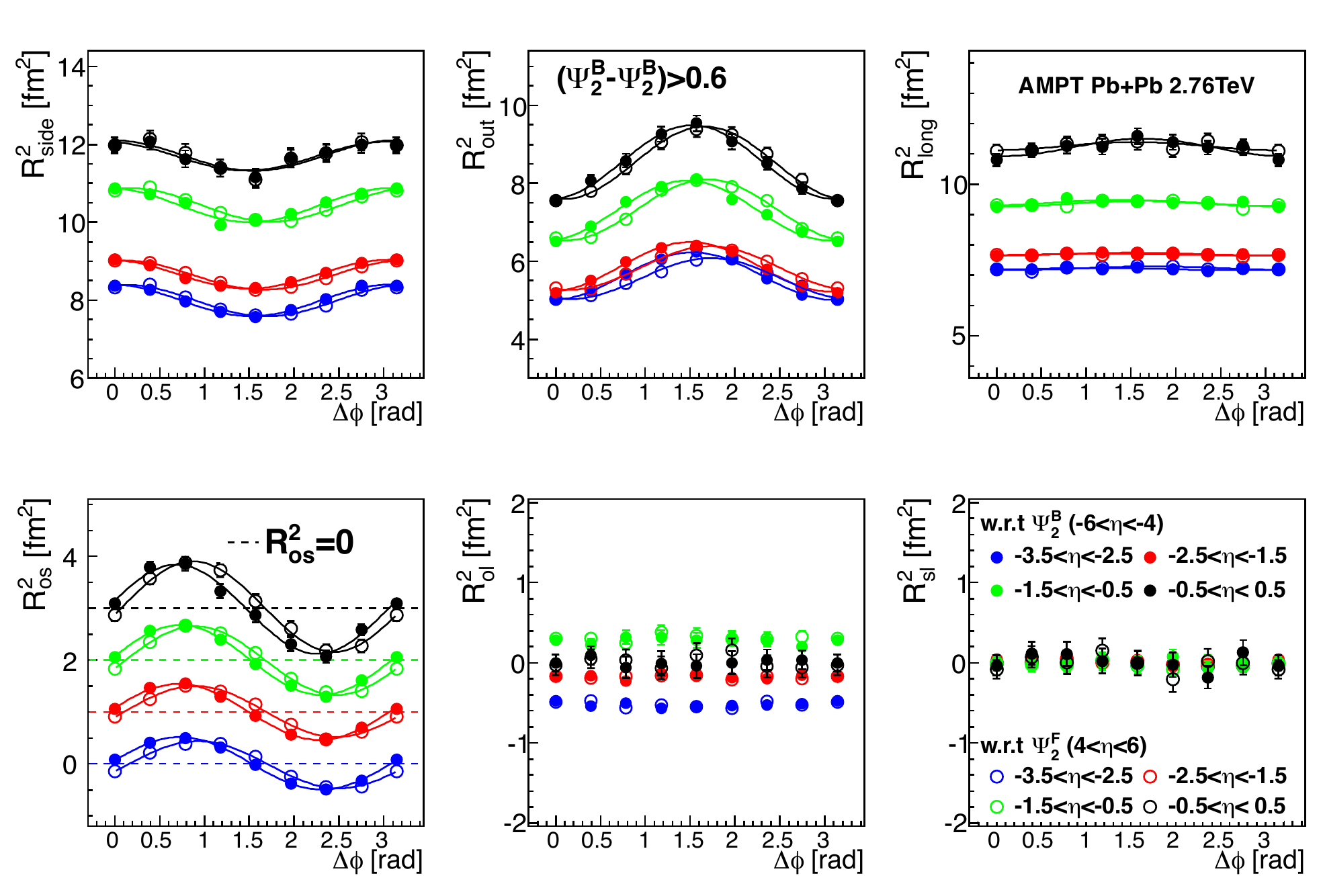}
\caption{\label{fig4} Azimuthal angle dependence of the radii parameters relative to the backward and forward $\Psi_2$ with the event cut of $(\Psi_2^{\rm B}-\Psi_2^{\rm F})>0.6$, where the dashed lines show $R_{\rm os}^2=0$.}
\end{center}
\end{figure}

Figure~\ref{fig3} shows $R_{\rm s}^2$, $R_{\rm o}^2$, and $R_{\rm os}^2$ as a function of azimuthal pair angle relative to the $\Psi_2^{\rm B}$ ($\Delta\phi$) for four $\eta$ regions, where $(\Psi_2^{\rm B}-\Psi_2^{\rm F})>0.6$ was required. The oscillations of the HBT radii measured in $\eta>0$ are shifted to the direction of $\Delta\phi<0$, which is the direction of $\Psi_2^{\rm F}$ in the current event cut. This phase shift can be understood to be a possible twist effect in the final source distribution.
The radii oscillations measured in $\eta<0$ have no phase shift because the negative $\eta$ is closer to the backward angle used for a reference of the HBT measurement and therefore the twist effect is supposed to be small.
%
%Fig. 3
%--------------------------------------------------------
%--------------------------------------------------------
%
Figure~\ref{fig4} shows all the extracted radii parameters measured with respect to $\Psi_{2}^{B}$ and $\Psi_{2}^{F}$ 
with the event cut of $(\Psi_{2}^{B}-\Psi_{2}^{F})>0.6$. The phase shift can be seen in the transverse radii (not in the longitudinal radii including the cross terms) when comparing the results for $\Psi_{2}^{B}$ and $\Psi_{2}^{F}$ in the same $\eta$ region. It indicates the twist of the event plane angles in the backward and forward angles. 
%
%Fig. 4
%--------------------------------------------------------
%--------------------------------------------------------
%
These oscillations were fitted with the following functions:
\begin{eqnarray}
R_{\mu}^2 (\Delta\phi) &=& R_{\mu,0}^2 + 2R_{\mu,2}^2 \cos(2 \Delta\phi + \alpha) \,\; ({\rm for} \; \mu=o, s),\\
R_{\mu}^2 (\Delta\phi) &=& R_{\mu,0}^2 + 2R_{\mu,2}^2 \sin( 2\Delta\phi + \alpha) \;\; ({\rm for} \; \mu=os),
\end{eqnarray}
to extract the magnitude of the phase shift ($\alpha$). The phase shift parameter $\alpha$ obtained from the results with respect to $\Psi_2^{\rm F}$ and $\Psi_2^{\rm B}$ is plotted as a function of $\eta$ in Fig.~\ref{fig4}. The $\alpha$ increases going from backward to forward angle in all cases. The variation of $\alpha$ in the $\eta$ dependence is comparable to the difference between results relative to $\Psi_2^{\rm F}$ and $\Psi_2^{\rm B}$ at the same $\eta$. The finite slope in the $\eta$ dependence of $\alpha$ indicates
the twisted source in the final state due to the initial longitudinal fluctuations, as well as the twisted event plane and flow as discussed in Ref.~\cite{ETW1,ETW2}.
In this study, the effect of the event plane resolution is assumed to be negligible, 
but there should be an uncertainty derived from the event cut of $(\Psi_2^{\rm B}-\Psi_2^{\rm F})$ and their finite event plane resolutions.
%
%Fig. 5
%--------------------------------------------------------
\begin{figure}[ht]
\begin{center}
\includegraphics[width=0.98\textwidth,clip,trim=0 0 0 0]{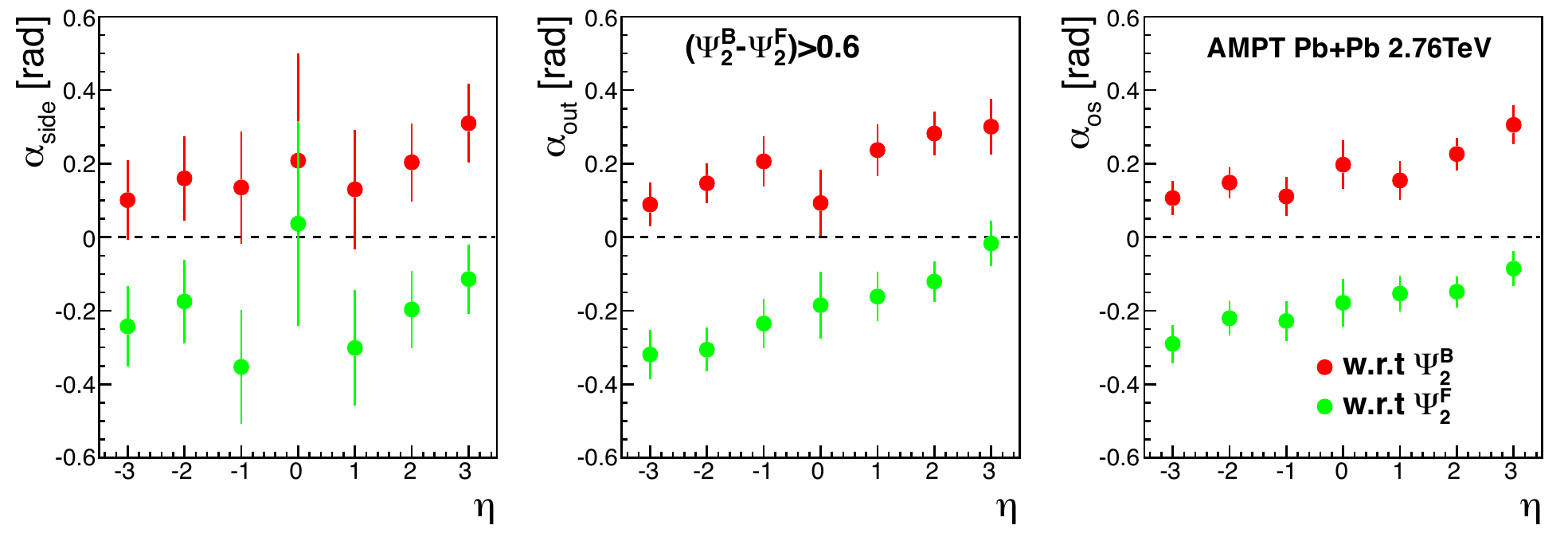}
\caption{\label{fig5} Phase shift parameters $\alpha$ obtained from the $R_{\rm s}^2$, $R_{\rm o}^2$, and $R_{\rm os}$ as a function of $\eta$. Results measured with respect to forward and backward event planes ($\Psi_2^{\rm F}$, $\Psi_2^{\rm B}$) are shown, with the event cut of $(\Psi_2^{\rm B}-\Psi_2^{\rm F})>0.6$.}
\end{center}
\end{figure}
%--------------------------------------------------------
%
\section{Summary}
We presented the results of HBT measurements using the event shape engineering for Au+Au collisions at $\sqrt{s_{_{\rm NN}}}=200$ GeV at the PHENIX experiment. We found that the higher $Q_2$ selection enhances the measured final source eccentricity as well as $v_2$. Although the model comparison is needed to disentangle both spatial and dynamical effects on the HBT radii, this study would clarify the relation between the initial and final source eccentricities and constrain better the system dynamics.

We also studied the event twist effect using AMPT model. When selecting events with finite difference between forward and backward event plane angles, the oscillations of HBT radii are shifted in the phase and the phase shift increases with $\eta$. The results indicate a possible twisted source in the final state preserving the initial twist due to the longitudinal fluctuations. This effect might be measured in experiments at RHIC and the LHC.

Both techniques could be useful to probe and control initial fluctuations in transverse plane and longitudinal directions, 
as well as to study the response of the system to the space-time evolution.

\end{document}